\documentclass[journal]{IEEEtran}

\usepackage{etex}
\usepackage{pifont}
\usepackage{bm}
\usepackage{cite}
\usepackage{amsmath,amssymb,amsfonts}
\usepackage{graphicx}
\usepackage{textcomp}
\usepackage[english]{babel}
\usepackage[normalem]{ulem}
\usepackage{graphicx,epstopdf}
\usepackage[colorlinks=true, allcolors=blue]{hyperref}
\usepackage[ruled,vlined]{algorithm2e}
\usepackage[table]{xcolor}
\usepackage{booktabs}
\usepackage{latexsym}
\usepackage{eurosym}
\usepackage{mathtools}
\usepackage{lscape}
\usepackage{xcolor}
\usepackage{lipsum}
\usepackage{optidef}
\usepackage{multirow}
\usepackage{siunitx}
\usepackage{subcaption}
\usepackage{enumitem}
\usepackage{orcidlink}
\usepackage{pdflscape}

\usepackage{amsthm}

\usepackage{tikz}

\usepackage{listings}
\usepackage{xcolor}
\begin{document}

\title{Semantic-Aware Adaptive Video Streaming Using Latent Diffusion Models for Wireless Networks}

\author{Zijiang~Yan$^{*}$\orcidlink{0000-0002-7959-8329},~\IEEEmembership{Graduate~Student~Member,~IEEE, }
        Jianhua~Pei$^{*}$\orcidlink{0000-0002-4066-9230}, 
         Hongda~Wu\orcidlink{0000-0001-8244-928X},~\IEEEmembership{Member,~IEEE, }
         
    Hina~Tabassum\orcidlink{0000-0002-7379-6949},~\IEEEmembership{Senior~Member,~IEEE,}
    and~Ping~Wang\orcidlink{0000-0002-1599-5480},~\IEEEmembership{Fellow,~IEEE}
\thanks{Z. Yan, J. Pei,  H. Wu, H. Tabassum and P. Wang are with the Department
of Electrical Engineering and Computer Science, York University, Canada. Z. Yan and J. Pei contributed equally to this work and are co-first authors.
}
}

\markboth{IEEE Wireless Communications,~Vol.~1, No.~1, October~2025}%
{Shell \MakeLowercase{\textit{et al.}}: Bare Demo of IEEEtran.cls for IEEE Journals}

\maketitle

\begin{abstract}
This paper proposes a novel {Semantic Communication (SemCom)}  framework for real-time adaptive-bitrate video streaming by integrating {Latent Diffusion Models} (LDMs) within the FFmpeg techniques. This solution addresses the challenges of high bandwidth usage, storage inefficiencies, and quality of experience (QoE) degradation associated with traditional {Constant Bitrate Streaming} (CBS) and {Adaptive Bitrate Streaming} (ABS). The proposed approach leverages LDMs to compress I-frames into a latent space, offering significant storage and semantic transmission savings without sacrificing high visual quality. {While retaining B-frames and P-frames as adjustment metadata to support efficient refinement of video reconstruction at the user side, the proposed framework further incorporates state-of-the-art denoising and Video Frame Interpolation (VFI) techniques.} 
These techniques mitigate semantic ambiguity and restore temporal coherence between frames, even in noisy wireless communication environments. Experimental results demonstrate the proposed method achieves high-quality video streaming with optimized bandwidth usage, outperforming state-of-the-art solutions in terms of QoE and resource efficiency. This work opens new possibilities for scalable real-time video streaming in 5G and future post-5G networks.
\end{abstract}


\raggedbottom
\IEEEpeerreviewmaketitle

\section{Introduction}

\IEEEPARstart{I}{n} today's digital age, the demand for streaming services with high-quality video content has surged alongside the proliferation of Internet-enabled devices, bridging the gap between media products and edge users \cite{zhou2022cadm}. Streaming services, which cater to a vast audience with diverse preferences, face the challenge of delivering superior video quality while optimizing wireless network bandwidth and resource usage. Real-time {Adaptive Bit-rate Streaming} (ABS) has emerged as a vital technology in this landscape, which enables dynamically adjusting video quality based on network conditions to provide a seamless viewing experience and improved {Quality of Experience (QoE)} \cite{bentaleb2024bitrate}. 
One significant advancement in ABS is the integration of {Fast Forward Moving Picture Experts Group} (FFmpeg), advanced video coding (H.264), {Hypertext Transfer Protocol (HTTP)}, and Dynamic Adaptive Streaming over HTTP (DASH). These open-source multimedia frameworks are capable of decoding, encoding, transmission, and streaming digital media files in various formats. In particular, FFmpeg’s flexibility and efficiency make it an indispensable tool for real-time video processing and ABS solutions.

{Compared with traditional video-streaming technologies, Semantic Communication (SemCom) provides a next-generation paradigm for more efficient, robust, and semantic-aware video transmission that addresses many limitations of conventional video streaming. Diffusion Models (DMs) \cite{blattmann2023stable}, a class of Artificial-Intelligence Generative-Content (AIGC) models, have demonstrated a remarkable ability to generate high-quality media. In a SemCom system, such models can be used for efficient encoding, robust denoising, and high-fidelity decoding, enabling intelligent fine-tuning of incoming video streams at the receiver and thereby improving both perceptual and semantic integrity of the reconstructed video frames. Fifth-generation (5G) and Beyond-5G (B5G) networks deliver unprecedented throughput and ultra-low latency, forming an ideal foundation for DM-enabled video streaming. However, these networks impose greater resource demands, introduce semantic ambiguities through encoding and decoding schemes that are highly sensitive to wireless channel noise, large-scale 4K and 6K streaming requirements, and the extensive use of machine-learning models, all of which must be managed to preserve end-to-end video quality. Fortunately, Latent Diffusion Models (LDMs) \cite{pei2024latent} have achieved excellent performance in wireless image transmission and video generation, and their advanced compression and channel-denoising techniques can outperform pixel-space DMs, ensuring seamless streaming with higher QoE and Quality of Service (QoS).}

{Inspired by both conventional FFmpeg video-streaming methods and emerging LDM-enabled SemCom for video transmission, we integrate LDM \cite{pei2024latent} into the FFmpeg framework to achieve high-quality real-time streaming that adapts dynamically to network fluctuations and diverse QoE requirements. Specifically, this paper first surveys the key challenges in conventional video transmission, SemCom-based approaches, and DM-enabled streaming. Next, we introduce an LDM-enabled semantic-aware ABS framework capable of delivering high-quality content over wireless channels by leveraging frame-type information\footnote{To efficiently transmit video within limited bandwidth, streaming standards decompose video into I-frames, P-frames, and B-frames \cite{pourreza2021extending}.}. The proposed scheme compresses I-frames into low-dimensional latent representations and encodes P and B frame motion as compact metadata, achieving significant bandwidth savings. At the receiver, LDMs denoise the semantic latent space of I-frames and refine the latent representations of P-frames and B-frames to enhance robustness against wireless channel noises. A Convolutional Neural Network–Gated Recurrent Unit (CNN–GRU)-based adaptive-bitrate controller further optimizes streaming decisions according to real-time network conditions, content characteristics, and user preferences.} We evaluate the system using a comprehensive QoE metric that balances video quality, quality fluctuations, and rebuffering risk, and demonstrate its superiority over state of the art ABR algorithms, including BOLA, Comyco, and MERINA \cite{kan2022improving}.

The rest of this paper is organized as follows. Section~\ref{problemformulation} introduces the existing challenges and limitations of conventional video streaming and SemCom for video transmission. {Section~\ref{subsection:relatedwork} provides a brief overview of DMs and LDMs, discusses SemCom-based video transmission methods built upon these models, and highlights the remaining open issues.} Section~\ref{proposedmethod} presents the design and workflow of the proposed LDM-enabled ABS framework, detailing its key components for compression, transmission, and video reconstruction. Section~\ref{casestudy} provides numerical results and performance analysis to validate the effectiveness of the proposed approach under real-time streaming scenarios. Finally, Section~\ref{conclusion} concludes the paper and outlines potential future research directions.

\section{Fundamental Challenges in SemCom for Video Streaming} \label{problemformulation}

{This section first outlines three major challenges in traditional video streaming: high bandwidth and storage consumption (Subsection~\ref{subsection:bandwidth}), low QoE (Subsection~\ref{subsection:lowQoE}), and Video Frame Interpolation (VFI) (Subsection~\ref{subsection:vfi}). It then introduces the concept of SemCom and how it addresses these challenges in Subsection~\ref{subsection:semcomforvs}. 
}

\subsection{High Bandwidth and Storage Utilization for CBS} \label{subsection:bandwidth}

\begin{figure}
\includegraphics[width=1\linewidth]{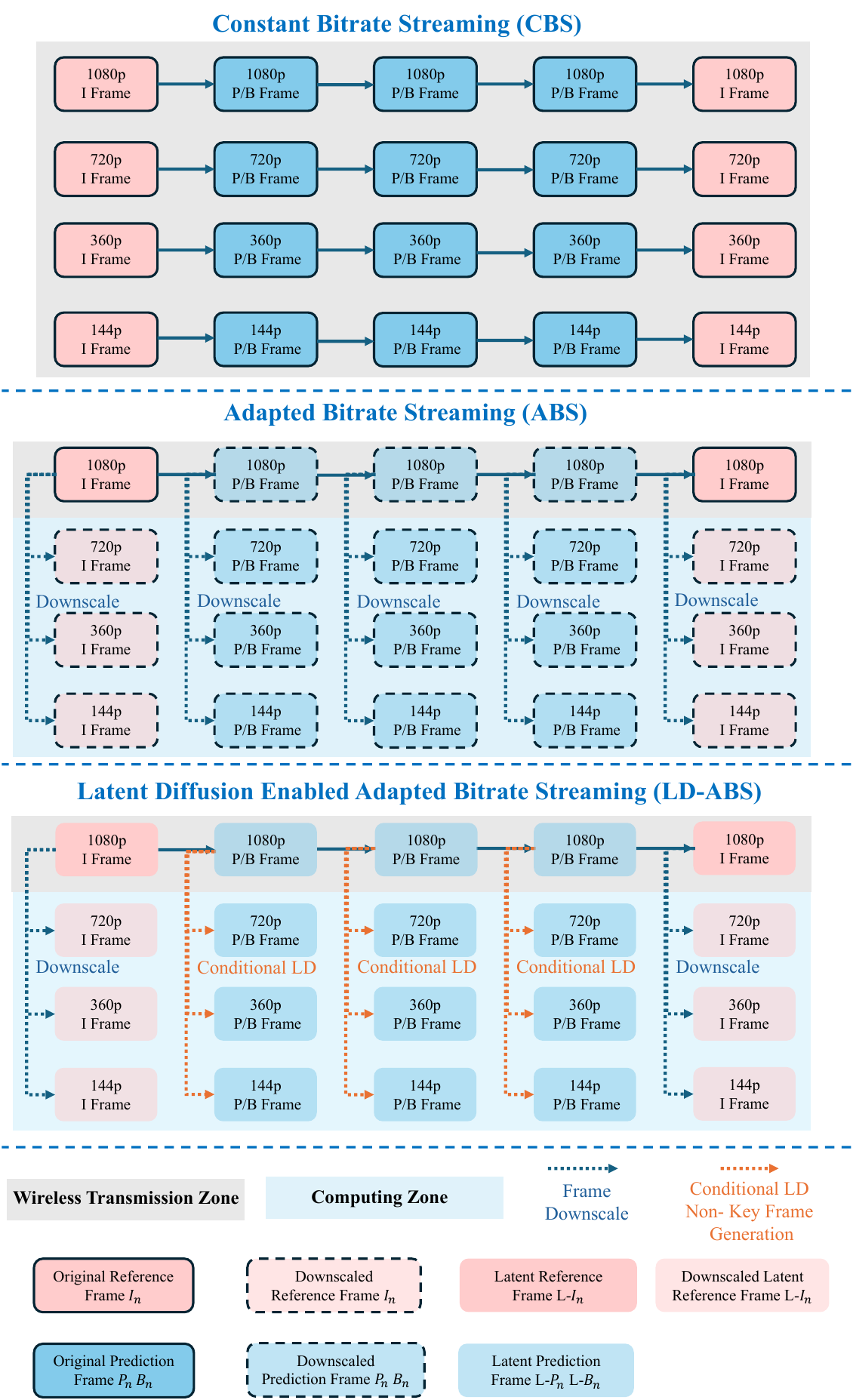}
\caption{Comparison of CBS, ABS and proposed LDM enabled ABS (LD-ABS) }
\label{fig:comparison_cbs_abs}
\end{figure}

Multi-bitrate streaming consumes substantial bandwidth and transmission resources. To meet diverse QoE demands, providers traditionally adopt {Constant Bitrate Streaming (CBS), as depicted in Fig.~\ref{fig:comparison_cbs_abs}}, wherein parallel streaming channels deliver videos at multiple resolutions and quality levels. The primary challenges associated with CBS are outlined as:

\textbf{(i) Bandwidth Consumption}: CBS requires allocating a fixed bandwidth for each resolution and quality level, regardless of the video's content complexity \cite{wang2023trafada}. This results in inefficient utilization of available bandwidth, especially during periods of low network congestion or when transmitting less complex scenes \cite{su2024reducing}. As shown in Fig.~\ref{fig:comparison_cbs_abs}, a typical CBS system maintains four parallel streams at 1080P, 720P, 360P, and 144P resolutions, respectively. Videos captured at 1080P are transcoded into different bitrates using FFmpeg, generating corresponding reference frames ($I_n$) and predictive frames ($P_n$ and $B_n$), where predictive frames are significantly smaller in size \cite{wiegand2003overview}. However, all frames across all resolutions must still be transmitted to users, as they manually select the desired resolution at the client side. This redundancy leads to substantial traffic and storage overhead, overburdening the {Content Delivery Network} (CDN) and causing excessive resource consumption.

\textbf{(ii) Server and Infrastructure Overload}: CBS places a heavy burden on servers, networking infrastructures, and cloud platforms. Managing numerous simultaneous requests for high-quality streams increases processing pressure, exacerbates network congestion, and leads to higher latency, frequent buffering, and even service disruptions during peak periods or large-scale events \cite{wang2023trafada, su2024reducing}. The need to sustain multiple concurrent streams for various resolutions further amplifies server-side storage and computation demands.

\textbf{(iii) Wireless Transmission Challenges}: In wireless communication environments, maintaining constant bitrates becomes highly challenging due to the inherent variability of signal quality and network conditions. High-bitrate streams under CBS rapidly deplete available wireless resources, leading to degraded QoE for mobile users, increased retransmissions, and higher traffic costs \cite{wang2023trafada}.

\subsection{Low QoE in Multi-Bitrate Adaptive Streaming}\label{subsection:lowQoE}

ABS dynamically adjusts video quality according to users’ network conditions and device capabilities, aiming to optimize viewing experiences while efficiently utilizing bandwidth. 
However, achieving high-quality video streaming without excessive bandwidth consumption remains challenging. Traditional ABS approaches often struggle to balance video quality and network efficiency, resulting in either compromised viewing experiences or unnecessary data consumption \cite{bentaleb2024bitrate}. The key challenges are summarized as follows:

\textbf{(i) Adaptation under Variable Wireless Conditions}:
ABS must dynamically adjust the video bitrate in response to fluctuating network environments. Frequent bitrate switching can introduce latency and instability, particularly in dynamic wireless scenarios. Under poor conditions, ABS tends to aggressively lower the bitrate, causing compression artifacts and degraded visual quality \cite{kan2022improving}, whereas CBS maintains a consistent quality level, albeit with limited flexibility.

\textbf{(ii) Cache and Buffer Management}:
Managing cached video chunks at multiple resolutions increases system complexity and storage demands \cite{wang2023trafada}. Efficiently prefetching, storing, and switching between segments remain a critical challenge for delivering seamless playback in ABS systems.

\textbf{(iii) Trade-off Between Smooth Playback and Quality Consistency}:
To ensure smooth and continuous playback, ABS frequently switches bitrates according to network variations. However, excessive bitrate fluctuation negatively impacts both QoE and QoS \cite{wang2023trafada}, leading to inconsistent viewing quality and user dissatisfaction.

\subsection{Video Frame Interpolation and Temporal Consistency}
\label{subsection:vfi}
Streaming video is typically composed of Intra-coded frames (I-frames), Predictive-coded frames (P-frames), and Bidirectionally predictive-coded frames (B-frames) \cite{pourreza2021extending}, defined as follows: 
\begin{itemize} 
\item \textbf{I-frames}: Independently compressed frames containing complete image information, similar to standalone images (e.g., PNG files). 
\item \textbf{P-frames}: Frames that predictively encode differences from preceding I-frames or P-frames using motion vectors and residual data. 
\item \textbf{B-frames}: Frames that predictively encode differences by referencing both preceding and following I-frames or P-frames, offering higher compression efficiency. 
\end{itemize}

Reconstructing video sequences from these frames is inherently challenging \cite{danier2024ldmvfi}. In video compression, I-frames serve as key references containing full image data, while P-frames and B-frames encode only motion vectors\footnote{Motion vectors represent object displacements between frames, enabling efficient compression by avoiding redundant storage of unchanged content.} and residual differences relative to reference frames. Although motion vectors are lightweight compared to full frames, enabling high compression ratios, accurate motion estimation between frames is difficult. Temporal consistency is particularly hard to maintain when using generative models for frame reconstruction \cite{danier2024ldmvfi}. Moreover, errors in motion estimation can cause misalignment and temporal artifacts, leading to QoE degradation  \cite{danier2024ldmvfi}.

\subsection{SemCom for Video Streaming and Challenges} 
\label{subsection:semcomforvs}
{SemCom focuses on transmitting the meaning of source data rather than raw bits or symbols. By extracting, encoding, and adapting semantic features, SemCom can directly address some challenges in traditional video transmissions, i.e.,}\\
{\textbf{(i) SemCom reduces bandwidth and storage:}
SemCom encoders extract only the essential semantic features of each video segment, eliminating parallel full-resolution streams and cutting redundant data. This feature-level compression sharply lowers both bandwidth use and storage requirements.}\\
{\textbf{(ii) Semantic-guided caching, buffering, and reconstruction:}
Edge or client caches store compact semantic descriptors instead of full video segments, simplifying buffer management and speeding startup. At playback, generative decoders use these semantic cues  to rebuild frames with strong temporal coherence and reduced interpolation artifacts.}


{Nevertheless, SemCom faces several challenges that motivate the use of AIGC models: \textbf{(i) Limited semantic reconstruction fidelity:} SemCom systems rely on light-weight models that struggle to reproduce fine details, complex textures, and rapid motions, resulting in blurring, artifacts, and temporal inconsistencies under demanding scenarios.} 
{\textbf{(ii) Wireless impairments and noise mitigation:}
Most SemCom system designs assume ideal or simplified channels and lack integrated denoising mechanisms, leaving semantic features vulnerable to real-world impairments such as multipath fading and interference.}

\section{State-of-the-Art: DM-Aided SemCom for Video Streaming} \label{subsection:relatedwork}
{To address the key challenges currently focused on in SemCom for video streaming, DMs and LDMs, as the most advanced AIGC models for visual data, offer effective solutions for high-quality data reconstruction and robust information transmission in SemCom. Therefore, this section first briefly reviews DMs and LDMs in Subsection \ref{subsection:overview}, then introduces several DM- and LDM-based approaches applied to video streaming in Subsection \ref{subsection:advance}, and finally summarizes the remaining limitations of these methods in Subsection \ref{subsection:challenges}.}

\subsection{Overview of DMs and LDMs}
\label{subsection:overview}

{DMs are a class of likelihood-based generative frameworks that learn to synthesize data by reversing a controlled noising process. During training, each data sample is gradually corrupted through a sequence of Gaussian noise steps; a neural network then learns to denoise these noisy samples step by step, ultimately recovering the original signal. While this pixel-space diffusion yields high-quality outputs, it demands substantial computation and memory, especially for high-resolution video.}
{Compared to DMs, LDMs address these challenges by shifting both the forward noise injection and reverse denoising into a lower-dimensional latent space. An encoder first projects each video frame into compact latent embeddings; diffusion and denoising occur entirely within this latent domain; a decoder then reconstructs the full-resolution frames \cite{pei2024latent}. Conditional LDMs (CLDMs) further enhance this process by integrating auxiliary information—such as motion vectors or semantic descriptors—into the denoising network. 
The key distinctions of LDMs compared to DMs for video streaming are:}

{ 
\begin{itemize}
 \item \textbf{Latent-space diffusion and dimensionality reduction:}
By encoding frames into a compact latent representation and performing diffusion steps in this lower-dimensional space, LDMs drastically reduce both memory usage and computational load, making high-resolution video generation more tractable.

\item \textbf{Multi-step denoising with temporal coherence:}
The sequential denoising process refines latent embeddings incrementally, enabling fine-grained reconstruction. This multi-step approach captures complex spatial details and temporal dependencies across adjacent frames, allowing robust recovery from noisy or incomplete latents.

\item \textbf{Conditioned Denoising:} 
By conditioning the denoising network on external cues (such as motion adjustment metadata, scene semantics, or object trajectories), CLDMs guide the generation toward temporally consistent and semantically accurate video frames even under varying network or content dynamics.
\end{itemize}
}

\subsection{Existing State-of-the-Art}
\label{subsection:advance}
\subsubsection{DM-aided SemCom for Video Streaming}
{

Zhou \textit{et al.} in \cite{zhou2022cadm} presented Codec-aware Diffusion Modeling (CaDM), a novel neural-enhanced video streaming paradigm. CaDM improves compression efficiency by reducing both the resolution and color bit-depth of video frames during encoding. At the decoder side, it employs a denoising diffusion process conditioned on the encoder's settings to restore high-quality frames, achieving significant bit-rate savings while maintaining superior visual quality.
Li \textit{et al.} in \cite{li2024extreme} proposed an extreme video compression approach leveraging the predictive power of the {Forward Diffusion Process} (FDP) with a pre-trained model.

}

\subsubsection{LDM-Aided SemCom for Video Streaming}
{
Danier \textit{et al.} in \cite{danier2024ldmvfi} presented LDMVFI, leveraging high-fidelity image synthesis capabilities of DMs for {Video Frame Interpolation (VFI)}. LDMVFI generates dynamic frames based on I-frames and incorporates a vector-quantized auto-encoding model, VQ-FIGAN, to enhance VFI performance.
Yu \textit{et al.} \cite{yu2024efficient} introduced Content-Motion LDM, which decomposes a video into a content frame (image-like) and a low-dimensional motion latent representation. 
Yu \textit{et al.} \cite{yu2023video} proposed the Projected Latent Video Diffusion Model (PVDM), consisting of an autoencoder stage and a DM stage. The autoencoder employs one latent vector to capture common content (e.g., background) while using two additional vectors to encode motion.
%
Ma \textit{et al.} in \cite{ma2025diffusion} presented DiffVC, a diffusion-based perceptual neural video compression framework that effectively integrates a foundational DM with the video conditional coding paradigm. This framework leverages temporal context from previously decoded frames and the reconstructed latent representation of the current frame to guide LDMs in generating high-quality images.

}

\subsection{Challenges}
\label{subsection:challenges}

{ 
Despite the significant potential demonstrated by DMs and LDMs in enhancing SemCom for video streaming, several challenges  persist, i.e.,
\begin{itemize}
\item \textbf{Computational cost and inference latency:}
Diffusion in pixel space requires dozens to hundreds of denoising steps per frame, resulting in prohibitively high inference latency and processing overhead \cite{li2024extreme}. Even when the entire diffusion process is moved into a lower-dimensional latent space, full-step denoising remains computationally intensive and cannot consistently meet the low-latency requirements of real-time video streaming.

\item \textbf{Underutilization of video coding structure (I/B/P-frames):}
Existing approaches \cite{danier2024ldmvfi, yu2024efficient, yu2023video, ma2025diffusion} primarily operate on consecutive frames. They often do not explicitly incorporate or leverage the traditional video coding structure involving I-frames and B/P-frames.

\item \textbf{Poor channel adaptability:}
Existing works encode and send video at a fixed frame rate and semantic bitrate, without accounting for real-world wireless channel fluctuations. Thus, they cannot maintain stable QoE or temporal consistency when network conditions vary, leading to buffering, frame drops, or inconsistent visual quality.
\end{itemize}

}

\section{Real-Time Latent Diffusion Adaptive-Bitrate Video Streaming Framework } \label{proposedmethod}

This section first presents a channel-aware bitrate selector that automatically determines the bitrate for the next video chunk based on channel conditions, semantic performance requirements, and user preferences, as described in Subsection~\ref{bitrateselector}. Next, the compression, denoising, and reconstruction components of the {Latent Diffusion Adaptive Bitrate video Streaming (LD-ABS)} framework are introduced for frame-by-frame video transmission at the selected resolution in Subsection~\ref{subframework}. Finally, Subsection~\ref{subsection:motivation} summarizes the motivations and contributions of the proposed framework.

\begin{figure*}
\includegraphics[width=1\linewidth]{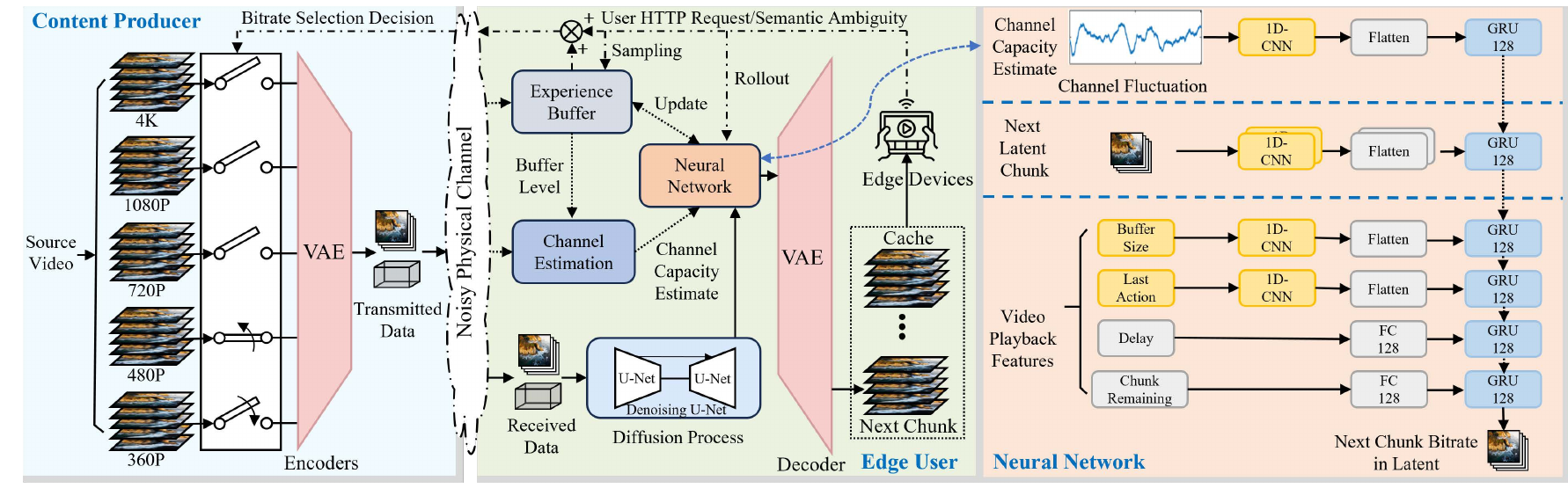}
\caption{ Proposed next chunk bitrate selection for adaptive bitrate streaming framework}
\label{fig:abr}
\end{figure*}

\subsection{Channel-Aware Video Streaming Bitrate Selector} \label{bitrateselector}

In practical scenarios, an edge user typically sends an HTTP GET request to request a streaming service. Overall, ABS utilizes CNN-GRU model to select the bitrate as the output of {Neural Networks} (NN) for the next chunk and the proposed ABS selector diagram is depicted in Fig. \ref{fig:abr}. We divide ABS selector into the following three sub-units, each corresponding to rendering streaming video \cite{huang2019comyco}:

\begin{itemize}
\item \textbf{Wireless Channel Resource Estimation:} 
This sub-unit estimates the variations of the noisy physical wireless channel. A channel state detector within the wireless communication network identifies the channel status and forwards this information to the neural network.

\item \textbf{Experience Buffer:} 
Inspired by offline Reinforcement Learning (RL) techniques, we use a buffer to store past expert strategies, allowing the algorithm to sample randomly from the buffer pool during the training process.

\item \textbf{Neural Network:} For each episode $e$, the ABS learning agent determines a suitable bitrate for the next chunk via a NN. The input features of this NN in $k$-th chunk are ($ S_k = \{ C_k, M_k, F_k \}$) and are detailed as follows: \textbf{\textit{1) Past Wireless Channel Estimate:}} The learning agent uses the past $t$ chunks' channel status vector $C_k = \{ c_{k-t}, \dots, c_k \}$ as input to the NN, where $c_i$ represents the channel throughput for video chunk $i$. \textbf{\textit{2) Latent Content:}} To detect the diversity of video content, the learning agent uses $M_k = \{ N_{k+1}, V_{k+1} \}$, where $N_{k+1}$ and $V_{k+1}$ denote the size of each bitrate of the next chunk $(k+1)$ and the perceptual quality metrics (semantic ambiguities) for each bitrate of the next chunk, respectively. \textbf{\textit{3) Player Rollout Playback:}} Rollout information is collected by the player on the edge user's side and is defined as $F_k = \{v_{k-1}, B_k, D_k, m_k\}$. Here, $v_{k-1}$ represents the video quality of the last video chunk selected, while $B_k$, $D_k$, and $m_k$ represent the buffer utilization,  loading time, and the normalized remaining chunks, respectively, for the past $t$ chunks.
\end{itemize}

Ultimately, the learning agent then selects the next chunk bitrate. The NN architecture includes a 1-dimensional CNN (1D-CNN), a Fully Connected 128-dimensional layer (FC-128), and GRUs that output 128-dimensional vectors (GRU-128). We use ReLU as the activation function and softmax for the last layer.  Consequently, as a part of LD-ABS, the ABS selector can mitigate the QoE loss challenges discussed in Section \ref{problemformulation} by selecting the optimal bitrate, thereby rationalizing bandwidth and storage consumption based on available resources, channel conditions, and QoS requirements.

\subsection{LDM-enabled Compression, Denoising, and Recontruction }
\label{subframework}

\begin{figure*}
\includegraphics[width=1\linewidth]{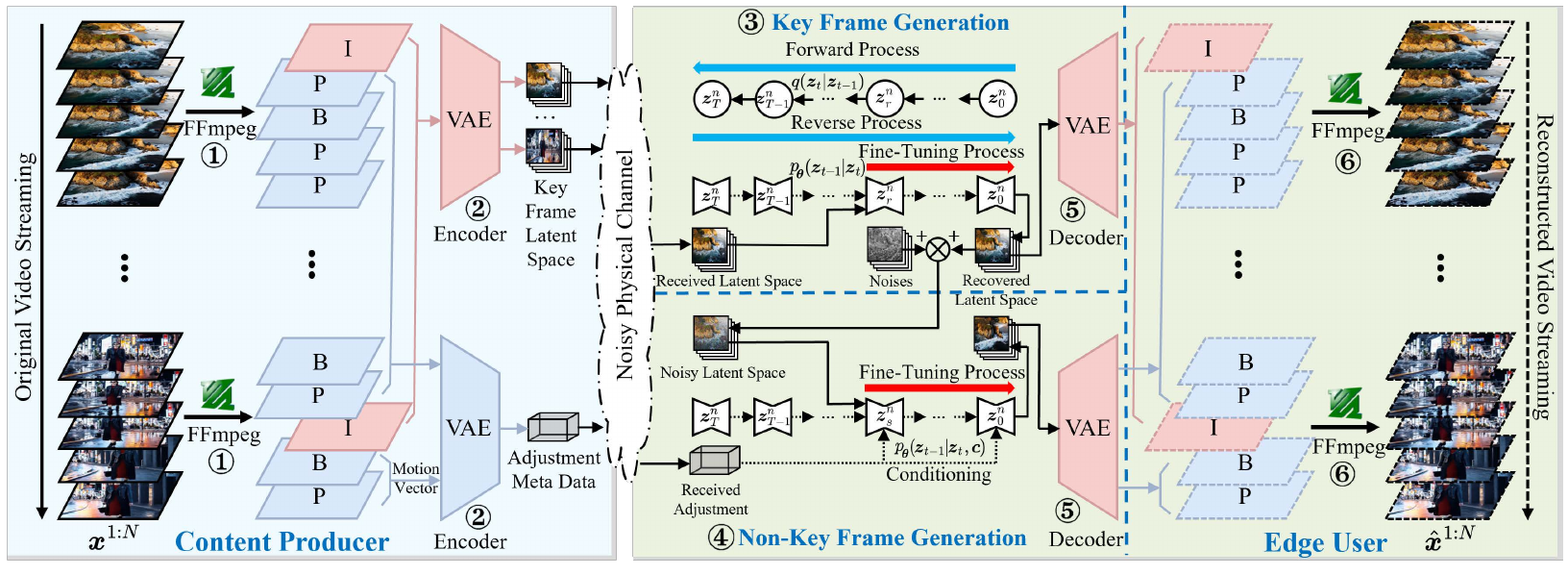}
\caption{ LDM-enabled adapted bitrate video streaming framework (LD-ABS)}
\label{fig:ldm-framework}
\vspace{-0.4cm}
\label{fig:architecture}
\end{figure*}

For the remainder of LD-ABS, to achieve higher compression of streaming video, we consider using LDM to compress the original reference frames into latent reference I frames. Meanwhile, we only retain the adjustment metadata for the predictive B and P frames. On the user side, latent reference frames are denoised to restore reference frames, and predictive frames are restored using conditional LDM and VFI. Recognizing the challenges faced by LDM-enabled real-time adaptive multi-band streaming, we propose a collaborative end-to-end LDM-enabled streaming framework. This framework aims to deliver high-resolution, low-latency streaming video while conserving communication bandwidth and storage with adaptive time-varying appropriate bitrate.
{
Overall, Fig. \ref{fig:ldm-framework} depicts the structure of the proposed LD-ABS framework , where $\bm{z}$ is the semantic latent features of video frames $\bm{x}$ encoded by VAE, $T$ represents the total number of forward steps of DM, $t$ denotes the $t$-th step, $n$ represents the $n$-th frame of the video stream with total $N$ frames, $\bm{\theta}$ is the parameters of the DM’s denoising U-Net neural network \cite{pei2024latent}, { which include 128 base channels in the base layer of the encoder with channel
multipliers [1, 2, 4, 8]}, $r$ represents the starting point for denoising received key frame semantic latent feature signals in the wireless channel, which is related to the channel state, $s$ represents the starting step for fine-tuning non-key frame semantic latent feature, and $\bm{c}$ is the condition for the conditional LDM, i.e., the small-size meta-data being transmitted. Furthermore, $q(\cdot|\cdot)$ denotes the forward process of the DMs, and $p_{\bm{\theta}}(\cdot|\cdot)$ represents the reverse denoising process.}
Specifically, the proposed framework involves the following key components:

\textbf{(i)} \textbf{Decode the video into I-frames, P-frames, and B-frames via FFmpeg} 

 In the beginning, We utilize FFmpeg to decode the streaming videos into their respective frame types in Fig. \ref{fig:architecture}, Step \scalebox{1.2}{\ding{172}}. It is crucial that each frame maintains accurate timestamps, which ensures that the streaming video can be reconstructed accurately in the final step. To achieve higher compression performance and reduce bandwidth usage, we do not compress each frame individually. Since I-frames are larger compared to other frames, we use the VAE in the LDM to compress I-frames into semantic latent features. B-frames and P-frames are left uncompressed, as compressing them would still consume significant computational resources. Moreover, these frame sizes are relatively small because they only store the differences from the previous or next frame.

\textbf{(ii) Semantic Encoding for I-Frames compression and B/P-Frames motion vector compression} 

In this phase, we utilize the VAE of LDM to compress I-frames, as depicted in Fig. \ref{fig:architecture}, Step \scalebox{1.2}{\ding{173}}. Since I-frames are comparatively larger than B-frames and P-frames, we compress I-frames to reduce bandwidth and storage requirements, thereby reducing the load on the CDN. The process involves using a key-frame VAE to compress I-frames into semantic Latent I-frames (L-I-frames) through the LDM process (Red VAE in Fig. \ref{fig:architecture}. On the other hand, B-frames and P-frames motion vectors and adjustments are encoded via the non-key-frame VAE (Blue VAE in Fig. \ref{fig:architecture}. To this end, it is no longer necessary to transmit the complete B/P-frames; instead, only the compressed L-I-frames and compressed motion vectors need to be transmitted, overcoming the bottleneck of high bandwidth and cache storage requirements discussed in Section \ref{problemformulation}.

\textbf{(iii) Transmitting semantic latent features  through noisy wireless channel}

As depicted in Fig. \ref{fig:architecture}, streaming video is sent to the edge user side via {Ultra-Reliable Low-Latency Communications }(URLLC). The fluctuating wireless environment provides an unstable communication bandwidth and channel degradation without prior awareness. ABS must first provide smooth, non-interruptive, low-latency service to satisfy QoS, and then aim to provide high bitrate streaming to satisfy QoE. With a constant {Channel-Bandwidth-Ratio} (CBR), the transmitter selects the appropriate bitrate based on the wireless environment, as different bitrates require distinct latent spaces to store latent frames. 

Furthermore, the 5G environments may introduce random channel noises and semantic errors during latent feature transmission processes, which can lead to incomplete or corrupted information delivery to the edge user side. As a result, not all frames in a streaming video are successfully or perfectly transmitted. Specifically, the transmission of L-I-frames, as well as the encoding information from B-frames and P-frames, may be affected by channel noises, attenuation and other uncertainties, further impacting the quality and integrity of the received video stream. Base Stations (BSs) and the proposed ABS mechanism process and schedule these requests, ensuring efficient resource allocation and maintaining QoE, considering computation complexity and streaming video quality.

\textbf{(iv) Restoring key frames through LDM denoising processes}

As highlighted in Fig.~\ref{fig:architecture}, Step \scalebox{1.2}{\ding{174}}, Upon receiving the L-I-frames $\bm{z}_r^n$, which contain wireless channel noises and errors, we apply a short reverse process to fine-tune the compressed L-I-frames $\bm{z}_r^n$ to restore the precise reference frames  $\bm{z}_0^n$. The fine-tuning step aims to improve the reconstruction quality by completing the channel denoising task. Finally, we pass $\bm{z}_0^n$ through the Red VAE decoder to reconstruct the I-frames. {Since the starting point $r$ for channel denoising adapts to the variations of accurately estimated Channel State Information (CSI), the proposed approach enhances the robustness of video transmission against challenges such as cahnnel noises, gain attenuation, and semantic errors under time-varying wireless network conditions.} Furthermore, since the LDM performs the denoising process in a low-dimensional semantic latent space with fewer steps, it achieves a relatively fast reconstruction speed. This helps mitigate the high computational complexity and slow processing typically associated with DMs.

\textbf{(v) Restoring non-key frames through conditional LDM} 

We follow this process to restore non-key frames. After recovering the I-frames $\bm{z}_0^n$, we apply a fine-tuning process based on the recovered latent space for L-I-frames as illustrated in Fig. \ref{fig:architecture}, Step \scalebox{1.2}{\ding{175}}. Specifically, a short forward noise $\bm{n}$ is added to the recovered latent space from the I-frames to determine $\bm{z}^n_s$, which prepares the content for reconstructing the B-frames and P-frames from the I-frames. Next, we apply a reverse diffusion process to denoise and restore the non-key latent features. This process is conditioned on the motion vector encodings of the B-frames and P-frames. Finally, the fine-tuned non-key latent features are decoded using the Red VAE decoder to generate Latent B-frames (L-B-frames) and Latent P-frames (L-P-frames). Since the semantic latent vectors of B/P-frames are fine-tuned based on similar L-I-frames, challenges commonly encountered in FFmpeg technologies, such as frame adjustment and motion estimation, are effectively addressed. Consequently, only the encodings of motion vectors, which serve as the condition $\bm{c}$ for the CLDM, need to be transmitted.

\textbf{(vi) Merging new frames to a new streaming video}

 The latent spaces of I-frames, B-frames, and P-frames are decoded by the key-frame VAE decoder as shown in Fig. \ref{fig:architecture} Step \scalebox{1.2}{\ding{176}}, to reconstruct the corresponding I-frames, B-frames, and P-frames. In the final step, as depicted in Fig. \ref{fig:architecture}, Step \scalebox{1.2}{\ding{177}}, we utilize FFmpeg to merge the newly constructed frames in chronological order to reconstruct the new streaming video.

\subsection{Key Benefits of LD-ABS} 
\label{subsection:motivation}

{

By combining adaptive bitrate selection with LDM-based compression and reconstruction tailored to different video frame types (I, P, B), LD-ABS provide following key benefits:

\begin{itemize}
\item \textbf{Enhanced Adaptability:} LD-ABS integrates bitrate selector that dynamically adapts streaming to varying network conditions and content resolutions.

\item \textbf{Reduced Computational Cost and Latency:} LD-ABS employs efficient LDMs and avoids lengthy reverse diffusion for every frame,  reducing computational load and latency compared to  DMs. A short reverse/denoising process is used for received I-frame latents (Fig.~\ref{fig:architecture}, Step \scalebox{1.2}{\ding{174}}), while a conditional LDM fine-tuning process reconstructs B/P-frames based on I-frame latents and motion data.


\item \textbf{Improved Bandwidth Efficiency and Robustness:} Bandwidth usage is reduced by transmitting only compressed I-frame latents (L-I-frames) and compressed motion vectors. Robustness against transmission errors is further strengthened by adaptive I-frame denoising based on channel estimates, while conditional reconstruction of B/P-frames ensures temporal coherence.
\end{itemize}

}

\section{Numerical Result and Discussion}
\label{casestudy}

\subsection{Experiment setup}
To assess performance of average chunk QoE, consistency, and fast adaptation across a wide range of wireless environments, we evaluate our proposed LD-ABS on a virtual player, which is widely assessing system performance  of video streaming  \cite{kan2022improving,huang2019comyco}. This player simulates the ABS process using real-world network throughput traces {(3rd Generation Partnership Project (3GPP), Federal Communications Commission (FCC), OBOE)} and allows comparison with other ABS algorithms under various user and network conditions \cite{kan2022improving}.
{Edge user-side simulations are performed on a desktop workstation equipped with an Intel Core i7-14700KF CPU, 32GB DDR4 RAM, and an NVIDIA RTX 4070 SUPER GPU (12GB VRAM).}

\subsection{Baseline Algorithms}
We evaluate LD-ABS against the following state-of-the-art ABS algorithms \cite{kan2022improving,huang2019comyco}:
\begin{itemize}
    \item \textbf{BOLA}: A buffer-based algorithm that employs Lyapunov optimization to select the optimal bitrate version while considering buffer occupancy constraints. 

    \item \textbf{RobustMPC}: A model-based algorithm that formulates the bitrate selection as an optimization problem over a future horizon of $h$ video chunks using model predictive control. The future throughput is estimated using the harmonic mean of the average throughput measured over the past five downloaded chunks. 

    \item \textbf{Comyco}: A model-free neural ABS algorithm that leverages neural networks to approximate an offline near-optimal solution through lifelong imitation learning. 

    { \item \textbf{Pensieve}: A Deep Reinforcement Learning (DRL)-based algorithm utilizing the A3C method to learn an optimal policy for rate adaptation by mapping buffer occupancy, throughput, and chunk size dynamics.}

\end{itemize}

\subsection{QoE Performance Comparison}

\begin{figure}[t]
    \centering

    \begin{subfigure}{1\linewidth}
        \centering
        \includegraphics[width=\linewidth]{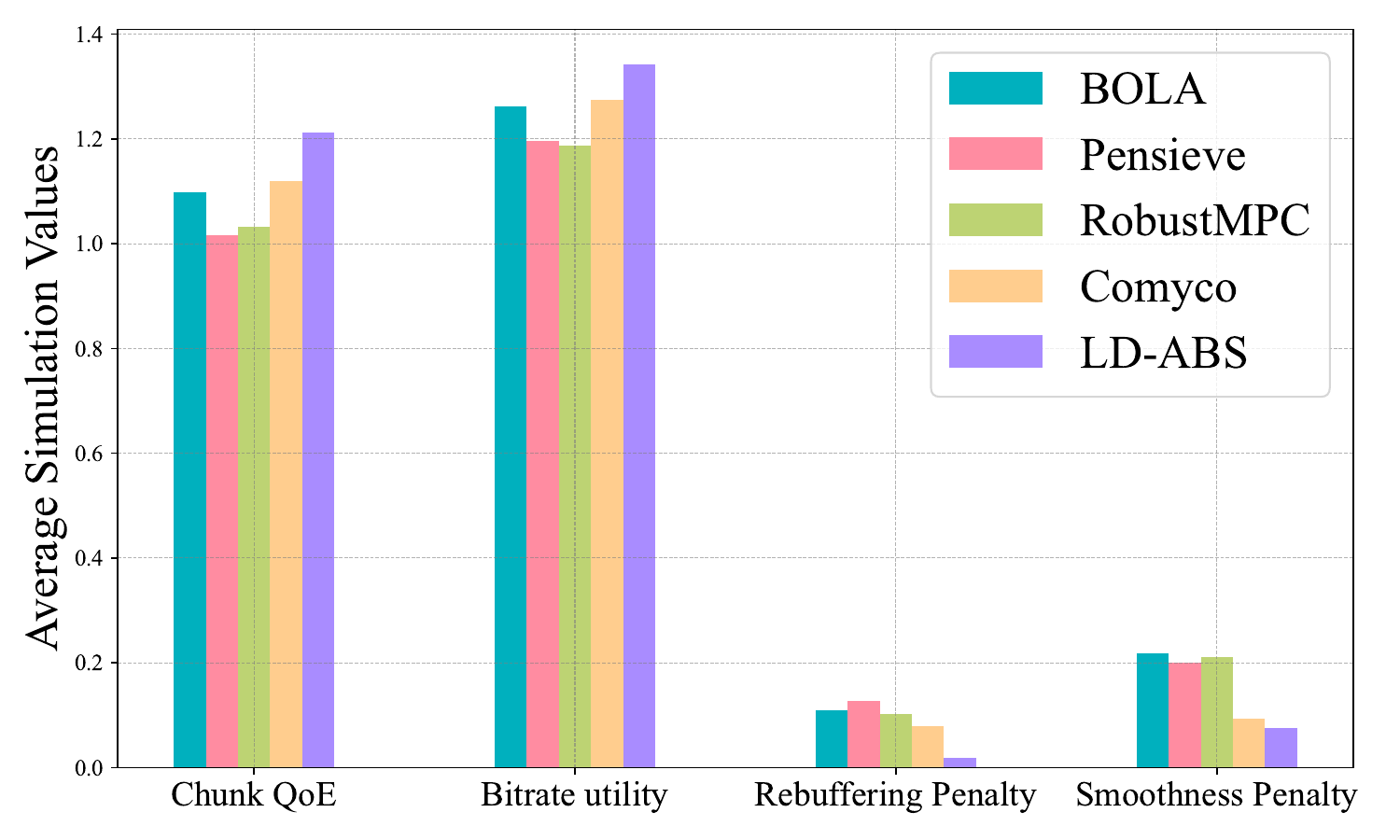}
        \caption{}
        \label{fig:combined_bars}
    \end{subfigure}
    
    \vspace{2mm} 

    \begin{subfigure}{1\linewidth}
        \centering
        \small
        {
        \begin{tabular}{l|c|c|c} 
            \toprule
            \textbf{Component} & \textbf{360p} & \textbf{720p} & \textbf{1080p} \\
            \midrule
            \multicolumn{4}{c}{\textbf{Transmitter Side}} \\
            \midrule
            I-frame extraction (\ding{172}) & 1.2 & 1.8 & 2.4 \\
            VAE encoding (\ding{173})         & 3.4 & 5.1 & 6.8 \\
            Z-frame compression               & 1.0 & 1.2 & 1.6 \\
            Metadata gen. (\ding{174})       & 0.7 & 1.0 & 1.3 \\
            \midrule
            \multicolumn{4}{c}{\textbf{Receiver Side}} \\
            \midrule
            Latent decoding (\ding{175})     & 3.6 & 5.5 & 6.9 \\
            Interpolation (\ding{176})       & 2.2 & 3.6 & 4.5 \\
            I-frame recon. (\ding{177})      & 2.8 & 4.4 & 5.9 \\
            \midrule
            \textbf{Total Latency}           & \textbf{14.9} & \textbf{22.6} & \textbf{29.4} \\
            \bottomrule
        \end{tabular}
        \caption{}
        \label{fig:latency_table}
        }
    \end{subfigure}

     \begin{subfigure}{1\linewidth}
        \centering
        \small
        {
        \begin{tabular}{l|c}
            \toprule
            \textbf{Streaming Method} & \textbf{End-to-End Latency} \\
            \midrule
            Traditional Broadcasting & 6,000 – 13,000 \\
            Pixel-Space DDPM         & 5,000 – 8,000  \\
            \textbf{LD-ABS}            & \textbf{2,000 – 3,000}  \\
            \bottomrule
        \end{tabular}
        \caption{}
        \label{tab:e2e_latency}
        }
    \end{subfigure}

    \caption{{(a) Performance evaluation of different ABS algorithms for FCC traces. (b) Latency breakdown for a chunk (in ms) of LD-ABS at different resolutions.  (c) Comparison of end-to-end average total streaming latency (in ms) across different methods.
}}
    \label{fig:combined_figure}
\end{figure}





\begin{figure*}[ht]
    \centering
    \scriptsize
    \includegraphics[width=1\linewidth]{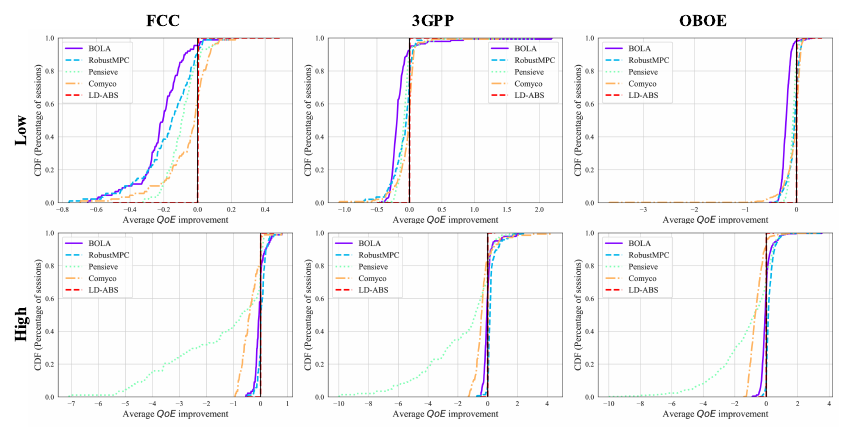}

    \caption{  {Cumulative Distribution Function} (CDF) of QoE for low-motion and high-motion scenarios across three trace types: FCC, 3GPP, OBOE}
    \label{fig:qoe_cdf_motion_traces}
\end{figure*}







 In ABS, the video is temporally divided into $K$ chunks (i.e., segments) of fixed duration $L$. Each chunk is encoded into multiple quality versions at different bitrates, with the set of available bitrates denoted as $\mathcal{A} = \{a_1, \dots, a_M\}$, where $M$ is the total number of bitrate options. $a_k$ represents the bitrate selected for the $k$-th chunk. 
 The Markov Decision Process (MDP) state $s_k \in \mathcal{S}$ for chunk $U_k$ is characterized by six features: (1)~previous chunk average throughput $C_{k-1}$, (2)~corresponding download time $d_{k-1}$, (3)~chunk sizes for all available bitrate versions of the $k$-th chunk, (4)~current buffer occupancy $B_{k-1}$, (5)~selected bitrate $a_{k-1}$ for the previous chunk, and (6)~the remaining number of chunks yet to be downloaded.
{To evaluate user-side QoE, we adopt an objective metric  that balances video quality (bitrate utility), quality fluctuations (smoothness penalty), and playback stalls (rebuffering penalty) through a linear combination \cite{kan2022improving}. The QoE for chunk \( k \) is defined as follows:
\begin{equation}
    \begin{aligned}
        \mathrm{QoE}(s_k, a_k) &= 
        m(a_k) - \alpha | m(a_k) - m(a_{k-1})| \\
        &\quad - \beta \max(0, d_k - B_{k-1})
    \end{aligned}
\end{equation}
where
\begin{itemize}
    \item \textbf{Bitrate Utility}: \( m(a_k) = \log(a_k/\min(\mathcal{A})) \) quantifies video quality perception, where higher bitrates improve visual quality but consume more bandwidth, increasing rebuffering risk.
    \item \textbf{Smoothness Penalty}: \( \alpha | m(a_k) - m(a_{k-1})| \) discourages large fluctuations in video quality between consecutive chunks, ensuring a consistent viewing experience.
    \item \textbf{Rebuffering Penalty}: \( \beta \max(0, d_k - B_{k-1}) \) penalizes playback stalls when the buffer runs out before the next chunk is downloaded.
\end{itemize}
The penalty weights are set as \( \alpha = 1 \) and \( \beta = 2.66 \) to balance smooth transitions and minimize rebuffering delays, following \cite{kan2022improving}.
The bitrate selection \( a_k \) for the \( k \)-th chunk is optimized to maximize the overall QoE based on the tradeoffs above.
}

 { The average CPU and GPU utilization for our edge users is 18\% and 41\%, respectively—values that are practical and achievable in real-world deployments.}
Fig.~\ref{fig:combined_bars} shows that our proposed LD-ABS framework outperforms the benchmarks presented in \cite{kan2022improving} in terms of chunk QoE, bitrate utility, rebuffering penalty, and smoothness penalty.
{ Fig.~\ref{fig:latency_table} illustrates that the latency for a sample video chunk is well within feasible limits across the processing pipeline shown in Fig.~\ref{fig:architecture}. Fig.~\ref{tab:e2e_latency} further demonstrates that LD-ABS reduces end-to-end latency compared to traditional approaches.}
 
\begin{figure}
    \centering
    \includegraphics[width=1\linewidth]{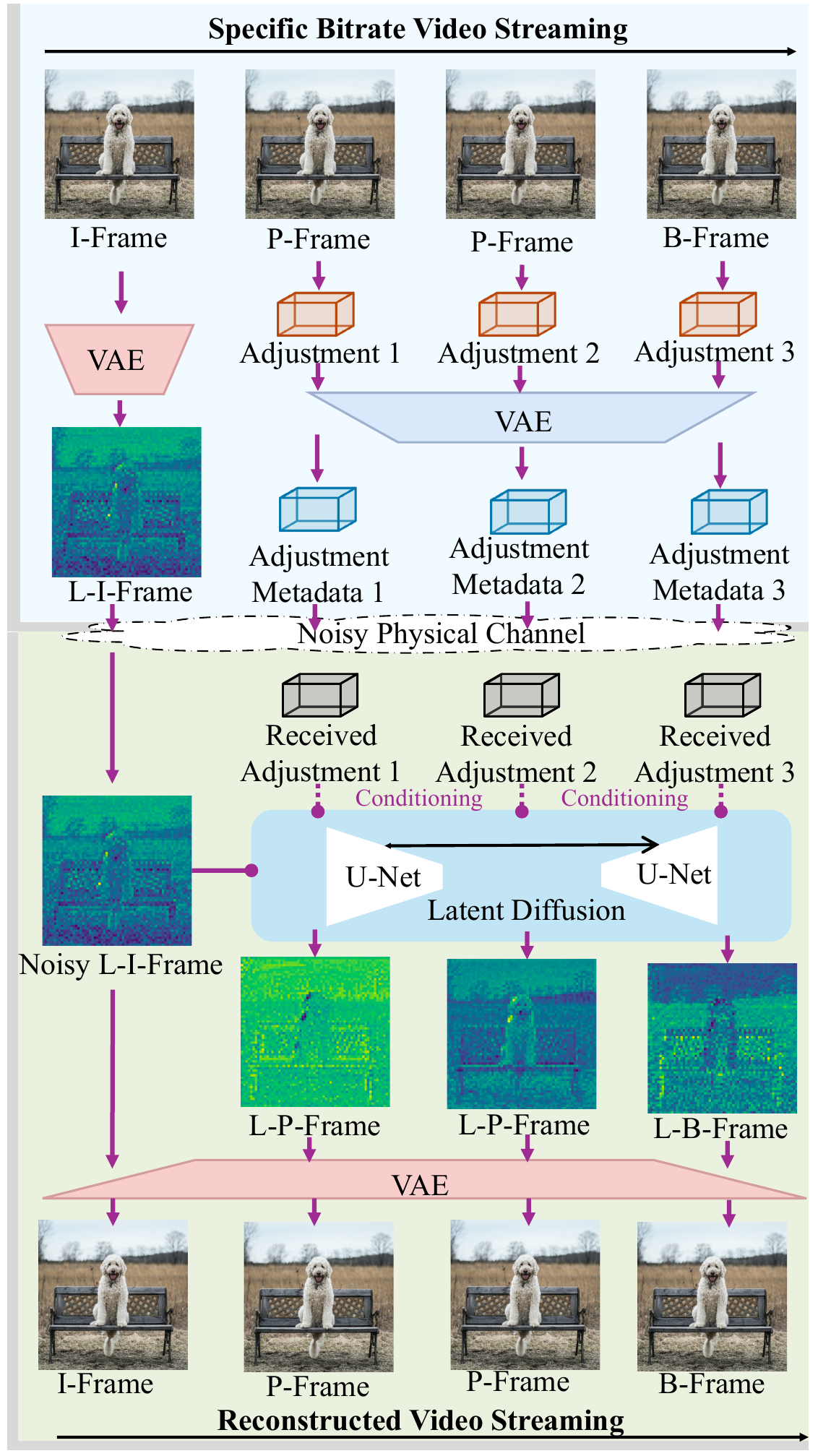}
    \caption{ Sample video chunk processing using the proposed LD-ABS framework.}
    \label{fig:demo}
\end{figure}

As depicted in Fig.~\ref{fig:qoe_cdf_motion_traces},
{LD-ABS outperforms the other benchmark with higher average QoE improvements across different traces. The other benchmark frameworks show comparable loss compared with our proposed LD-ABS. With CSI error, LD-ABS loses \textless 3 \% QoE, four times less than the closest competitor.  
The reason is two-fold: (i) denoising is performed in a low-dimensional latent space with only a handful of reverse steps, so the error does not propagate deeply; (ii) the subsequent conditional diffusion for B/P-frames is {anchored} to the already-refined I-frame, effectively absorbing residual noise. 
}
{Fig.~\ref{fig:demo} illustrates a sample streaming simulation within a chunk. This chunk consists of one I-frame followed by a P-frame, another P-frame, and a B-frame. Instead of transmitting the entire frames through a noisy wireless channel, only lightweight adjustment metadata is transmitted. The LDM then reconstructs the respective latent frames based on the recovered latent I frame ($z_o^n$) and the received adjustment metadata, generating the corresponding adjusted frames accordingly.
 }

\section{Conclusion and Future Directions} \label{conclusion}

This paper proposes a novel LDM-aided adaptive bitrate streaming framework to improve the efficiency and quality of real-time video streaming over wireless networks. The LDM-based approach could significantly reduce bandwidth consumption with high perceptual quality by semantic-aware video compression and reconstruction. It combines diffusion-based denoising with VFI to enhance temporal coherence and reduce the impact of noisy and attenuated wireless channels. Our experimental results confirm that the LD-ABS surpasses existing ABS methods in terms of QoE, robustness across sessions, and network adaptability.

While the above results look promising, many challenges are still open for future research. The multi-step reverse diffusion process introduces some latency in reconstructing the video, in particular for longer video sequences. Investigating acceleration methods like knowledge distillation and model quantization is imperative. Second, since current LDMs are only trained within a specific dataset, they might lack the generalization ability across diverse video contents and network conditions. Future work should be devoted to domain adaptation and continual learning to enhance the robustness in real-world streaming scenarios, such as live news broadcasts and sports streaming, as well as vehicular networks.

\bibliographystyle{IEEEtran}
\bibliography{ref_clean.bib}

\begin{IEEEbiographynophoto}{Zijiang Yan} [S] (zijiang.yan@ieee.org)
received the B.S. degree with a double major in Computer Science and Statistics from York University, Toronto, ON, Canada, in 2021. He is currently a Research Assistant with the Department of Electrical Engineering and Computer Science, Lassonde School of Engineering, York University.
His research interests include AI-enabled communication, Quantum Machine Learning, Diffusion Models, Large Language Models. He received 3rd place in the 2025 Student Innovation Competition on Sustainable Space Communications, hosted by the Satellite and Space Communications Committee (SSC) of the IEEE Communications Society (\textsc{IEEE ComSoc}). He also won the Lassonde Undergraduate Research Award (LURA) from York University in 2021.
\end{IEEEbiographynophoto}

\begin{IEEEbiographynophoto}{Jianhua Pei} [M] (jianhuapei98@gmail.com) received the B.Eng. and
      Ph.D. degrees in electrical engineering from Huazhong University of Science and Technology
      (HUST), Wuhan, China, in 2019 and 2025, respectively. He is currently working at the Central
      China Branch of State Grid Corporation of China. He was also a visiting Ph.D. student at the
      Department of Electrical Engineering and Computer Science, Lassonde School of Engineering, York
      University, Canada, in 2024. His research interests include power system data quality
      improvement, power system dynamics, power system cybersecurity, and artificial intelligence
      applications for communications.
\end{IEEEbiographynophoto}
    
\begin{IEEEbiographynophoto}{Hongda Wu} [M] (h.wu.yorku@gmail.com)
    received his Ph.D. degree in Electrical Engineering \& Computer Science from York University,
      Canada, in 2024. His research interests include federated learning, reinforcement learning,
      wireless networks, and the Internet of Things. He won the Best Paper Award from the \textsc{IEEE
      Wireless Communications and Networking Conference (WCNC)} in 2022.
\end{IEEEbiographynophoto}

 \begin{IEEEbiographynophoto}{Hina Tabassum} [SM] (hinat@yorku.ca)
   received the Ph.D. degree from the King Abdullah University of  Science and Technology, Thuwal, Saudi Arabia. She  was appointed as a Visiting Faculty with University of Toronto, Toronto, ON,
      Canada, in 2024, and the York Research Chair of 5G/6G-enabled mobility and  sensing applications
      in 2023, for five years. She is  currently an Associate Professor with the Lassonde  School of
      Engineering, York University, Toronto,  where she joined as an Assistant Professor, in 2018.
      She has co-authored more than 120 refereed articles in well-reputed \textsc{IEEE} journals,
      magazines, and conferences. Her current research interests include multi-band 6G wireless
      communications and sensing networks,  connected and autonomous systems, AI-enabled network
      mobility, and resource management solutions. She has been selected as the \textsc{IEEE ComSoc}
      Distinguished Lecturer for the term 2025-2026. She is listed in the Stanford’s list of  the
      World’s Top Two-Percent Researchers in 2021-2024. She was the recipient of the Lassonde
      Innovation Early-Career Researcher Award in 2023 and the \textsc{N2Women}: Rising Stars in
      Computer Networking and Communications in 2022.
\end{IEEEbiographynophoto}
   
\begin{IEEEbiographynophoto}{Ping Wang} [F] (pingw@yorku.ca)
   is a Professor at the Department of Electrical Engineering and Computer Science, York University, and a Tier 2 York Research Chair. Prior to that, she was with Nanyang Technological University, Singapore, from 2008 to 2018. Her recent research interests focus on integrating Artificial Intelligence (AI) techniques into communications networks. Her scholarly works have been widely disseminated through top-ranked \textsc{IEEE} journals/ conferences and received the \textsc{IEEE Communications Society} Best Survey Paper Award in 2023, and the Best Paper Awards from \textsc{IEEE} prestigious conference \textsc{WCNC} in 2012, 2020 and 2022, from \textsc{IEEE Communication Society: Green Communications \& Computing Technical Committee} in 2018, from \textsc{IEEE} flagship conference \textsc{ICC} in 2007. She has been serving as the associate editor-in-chief for \textsc{IEEE Communications Surveys \& Tutorials} and an editor for several reputed journals, including \textsc{IEEE Transactions on Wireless Communications} and \textsc{IEEE Transactions on Network Science and Engineering}. She is a Fellow of the \textsc{IEEE} and a Distinguished Lecturer of the \textsc{IEEE Vehicular Technology Society (VTS)}. She is also the Chair of the Education Committee of \textsc{IEEE VTS}.
\end{IEEEbiographynophoto}
\end{document}